\documentclass[sigplan,screen]{acmart}
\copyrightyear{2024}
\acmYear{2024}
\setcopyright{acmlicensed}\acmConference[ASPLOS '24]{29th ACM International Conference on Architectural Support for Programming Languages and Operating Systems, Volume 2}{April 27-May 1, 2024}{La Jolla, CA, USA}
\acmBooktitle{29th ACM International Conference on Architectural Support for Programming Languages and Operating Systems, Volume 2 (ASPLOS '24), April 27-May 1, 2024, La Jolla, CA, USA}
\acmDOI{10.1145/3620665.3640423}
\acmISBN{979-8-4007-0385-0/24/04}
\newcommand{\asplossubmissionnumber}{612}

\newcommand{\proj}{\textsc{GMLake}}
\usepackage{lipsum}
\usepackage[normalem]{ulem}
\usepackage{tikz}
\usepackage{graphicx}
\usepackage{subcaption}
\usepackage{caption}
\usepackage{algorithmic}
\usepackage[ruled,vlined,linesnumbered,algo2e]{algorithm2e}
\usepackage{amsmath}
\usepackage{ulem}

\newcommand{\Fig}[1]{Figure~\ref{#1}}

\newcommand{\Tbl}[1]{Table~\ref{#1}}
\newcommand{\Sec}[1]{Section~\ref{#1}}
\newcommand{\fixme}[1]{{\textcolor{black}{{#1}}}}

\newcommand\circlednumber[1]{%
  \begin{tikzpicture}[baseline=(char.base)]
    \node[shape=circle,draw,fill=black,inner sep=1pt] (char) {\textcolor{white}{\scriptsize\sffamily\bfseries#1}};
  \end{tikzpicture}}

\title[\proj{}: Efficient and Transparent GPU Memory Defragmentation ...]{\proj{}: Efficient and Transparent GPU Memory Defragmentation for Large-scale DNN Training \\ with Virtual Memory Stitching}

    \author{Cong Guo} 
    \affiliation{
        \institution{Shanghai Jiao Tong University} 
        \country{}
        }
    \affiliation{
        \institution{Shanghai Qi Zhi Institute}
        \country{}
    }
    \authornote{Equal contribution}

    \author{Rui Zhang}
    \affiliation{
        \institution{Ant Group}
        \country{}
    }
    \authornotemark[1]

    \author{Jiale Xu} 
    \affiliation{
        \institution{Shanghai Jiao Tong University} 
        \country{}
        }
    \affiliation{
        \institution{Shanghai Qi Zhi Institute}
        \country{}
    }
    
    \author{Jingwen Leng}
    \affiliation{
        \institution{Shanghai Jiao Tong University} 
        \country{}
        }
    \affiliation{
        \institution{Shanghai Qi Zhi Institute}
        \country{}
    }
    \authornote{Corresponding authors}

    \author{Zihan Liu} 
    \affiliation{
        \institution{Shanghai Jiao Tong University} 
        \country{}
        }
    \affiliation{
        \institution{Shanghai Qi Zhi Institute}
        \country{}
    }

    \author{Ziyu Huang}  
    \affiliation{
        \institution{Shanghai Jiao Tong University} 
        \country{}
        }
    \affiliation{
        \institution{Shanghai Qi Zhi Institute}
        \country{}
    }
    
    \author{Minyi Guo} 
    \affiliation{
        \institution{Shanghai Jiao Tong University} 
        \country{}
        }
    \affiliation{
        \institution{Shanghai Qi Zhi Institute}
        \country{}
    }
    \authornotemark[2]

    \author{Hao Wu} 
    \affiliation{
        \institution{Ant Group}
        \country{}
    }
   
    \author{Shouren Zhao} 
    \affiliation{
        \institution{Ant Group}
        \country{}
    } 

    \author{Junping Zhao} 
    \affiliation{
        \institution{Ant Group}
        \country{}
    }
    \authornotemark[2]
 
    \author{Ke Zhang} 
    \affiliation{
        \institution{Ant Group}
        \country{}
    }


\date{}

\begin{abstract}
Large-scale deep neural networks (DNNs), such as large language models (LLMs), have revolutionized the artificial intelligence (AI) field and become increasingly popular.
However, training or fine-tuning such models requires substantial computational power and resources, where the memory capacity of a single acceleration device like a GPU is one of the most important bottlenecks. 
Owing to the prohibitively large overhead (e.g., $10 \times$) of GPUs' native memory allocator, DNN frameworks like PyTorch and TensorFlow adopt a caching allocator that maintains a memory pool with a splitting mechanism for fast memory (de)allocation.
Unfortunately, the caching allocator's efficiency degrades quickly for popular memory reduction techniques such as recomputation, offloading, distributed training, and low-rank adaptation.
The primary reason is that those memory reduction techniques introduce frequent and irregular memory (de)allocation requests, leading to severe fragmentation problems for the splitting-based caching allocator.  
To mitigate this fragmentation problem, we propose a novel memory allocation framework based on low-level GPU virtual memory management called GPU memory lake (\textbf{\proj{}}).
\proj{} employs a novel \textbf{virtual memory stitching} (VMS) mechanism, which can fuse or combine non-contiguous memory blocks with a virtual memory address mapping. 
\fixme{\proj{} can reduce average of 9.2~GB (up to 25~GB) GPU memory usage and 15\% (up to 33\% ) fragmentation among eight LLM models on GPU A100 with 80~GB memory.
}
\proj{} is completely transparent to the DNN models and memory reduction techniques and ensures the seamless execution of resource-intensive deep-learning tasks. \fixme{We have opensourced \proj{} at }\href{https://github.com/intelligent-machine-learning/glake/tree/main/GMLake}{https://github.com/intelligent-machine-learning/glake/tree/main/GMLake}.
\end{abstract}

\begin{document}

\maketitle

\renewcommand{\shortauthors}{C.Guo, R.Zhang et al.}

\thispagestyle{empty}

\section{Introduction}\label{sec:introduction}

Large-scale deep neural network (DNN) models, specifically Large Language Models (LLMs), have revolutionized natural language processing (NLP) and artificial intelligence (AI) research~\cite{zhao2023survey}. LLMs, such as the GPT-3~\cite{GPT3} architecture, are sophisticated DNN models with remarkable capabilities in understanding, generating, and processing human language. These models leverage vast amounts of textual data and employ Transformer-based architectures characterized by attention mechanisms~\cite{vaswani2023attention} to achieve state-of-the-art performance on various NLP tasks. 
However, the widespread adoption of LLMs comes with significant computational challenges, as training or fine-tuning such models requires substantial computational power and resources. 
For example, OPT~\cite{zhang2022opt}, with 175 billion parameters, needs 34 days on 1,024 A100 GPUs, and 65B-parameter LLaMA~\cite{touvron2023llama} processes 1.4T tokens on 2048 A100 GPUs taking approximately 21 days.

Therefore, deep learning (DL) frameworks, e.g., PyTorch \cite{pytorch2018pytorch} and TensorFlow~\cite{tensorflow2022}, have emerged as the fundamental infrastructure for DNN models due to \fixme{their} flexibility and computational efficiency. 
Those DL frameworks have enabled the training of increasingly large and complex neural network models. 
Meanwhile, the GPU architecture~\cite{khalilov2021performance, choquette2020nvidia, otterness2020amd} \fixme{has} become the most widely used hardware to support the high-performance execution of DNN models.
On the other side, the growing scale and complexity of DNN models poses new challenges to GPU memory management.  
For instance, using the CUDA's native memory allocation APIs like \texttt{cudaMalloc} and \texttt{cudaFree}  incurs a large overhead.
To improve the efficiency of GPU memory allocation, DL frameworks opt to implement a caching allocator that maintains a memory pool with the best fit with coalescing (BFC) algorithm~\cite{tensorflow2022}.
Our experiments show that the caching allocator outperforms the native memory allocator by almost $10\times$.

On the other side, the rapid growth in the memory requirements of large-scale DNN models~\cite{sun2019gradientflow, ahn2022ai} has sparked the development of methods at the system- and algorithm-level to alleviate memory demands. 
Examples for these methods include recomputation~\cite{kusumoto2019graph, yang2019recomputation}, offloading~\cite{ren2021zero}, distributed training~\cite{lepikhin2020gshard, shoeybi2019megatron, xu2023efficient, harlap2018pipedream, huang2019gpipe, li2021chimera} and low-rank adaptation~\cite{hu2021lora}.
Even though these optimizations can effectively reduce memory footprint for training or fine-tuning large-scale DNN models, they may lead to poor memory utilization.
The reason is that they also introduce a significant amount of regularity and dynamicity in the memory allocation requests, which results in up to 30\% GPU memory fragmentation.

\begin{figure}[t]
    \centering
    \includegraphics[width=0.48\textwidth]{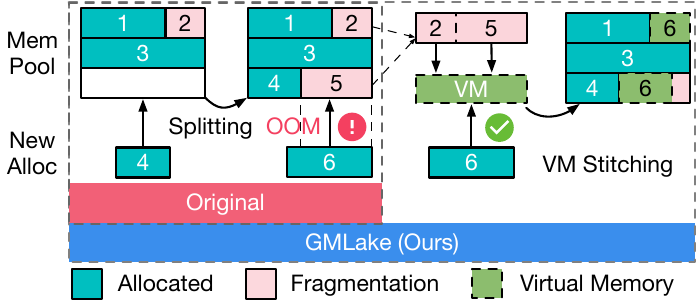}
    \caption{Representative example of memory allocation problem. The original splitting method can boost GPU memory utilization but cause fragmentation. Our proposed virtual memory stitching can complement and optimize the memory fragmentation issues.}
    \label{fig:intro}
\end{figure}

As shown in \Fig{fig:intro} left, DL frameworks manage the memory allocation within the memory pool.
They adopt the ``splitting'' method to split the memory pool to fit the DNN tensors' arbitrary size and boost the utilization of the memory pool. 
However, that will cause severe memory fragmentation for some new allocations. 
For example, the framework splits the third line to store the new allocation of Block 4. But the memory pool cannot hold Block 6 because the size of Block 6 is larger than Block 5, which cannot be exploited and becomes fragmented.
Finally, the framework will report the out-of-memory (OOM) error, one of the most common issues in the training processing for DNN models.
The aforementioned memory reduction techniques like recomputation and offloading can mitigate the OOM issue, but also lead to more frequent and irregular memory allocation and deallocation requests, exacerbating the fragmentation problem.

To mitigate GPU memory fragmentation and improve efficient memory utilization, this study focuses on exploring the causes of GPU memory fragmentation and proposes a novel memory allocation framework based on low-level GPU virtual memory management, called GPU memory lake (\textbf{\proj{}}),  to optimize GPU memory management with low overhead. As shown in \Fig{fig:intro} right, \proj{} employs a novel \textbf{virtual memory stitching} (VMS) mechanism, which is seemingly a reverse behavior to the splitting. 
Compared to the original framework, it can fuse or combine non-contiguous memory blocks with a virtual memory address mapping. 
For example, the VMS can map Block 6 to the Block 2 and 5 stitched block by a virtual memory address, then store Block 6 in the physical memory chunks of Block 2 and 5.
Obviously, virtual memory stitching effectively reduces memory fragmentation and improves memory utilization.
We implement the \proj{} on the low level of the DL framework and replace the original memory allocation API of DNN training.
\proj{} is completely transparent to the DNN models and other memory optimization methods, e.g., recomputation and offloading, ensuring the seamless execution of resource-intensive deep-learning tasks.

Overall, this work makes the following contributions:
\begin{itemize}
    \item We perform a characterization study to show that the caching allocator used in existing DL frameworks \fixme{suffers} from up to 30\% memory fragmentation when running large-scale DNN models with various memory reduction techniques such as recomputation, offloading, distributed training, and low-rank adaptation.
	\item We design and implement \proj{}, a novel memory allocator that effectively reduces memory fragmentation and improves memory utilization. \proj{} is transparent to the above existing memory reduction techniques. It integrates the virtual memory stitching mechanism by using the low-level CUDA virtual memory management knobs.
	\item We evaluate \proj{} on multiple prominent LLM optimization platforms with a set of representative open-source LLMs to demonstrate its effectiveness, efficiency, and robustness. In the best case, we can reduce GPU memory usage by \fixme{33\%}, translating to \fixme{25~GB} memory saving \fixme{comparing to native caching allocator in PyTorch} on an A100 GPU with 80~GB HBM memory.
\end{itemize}

\begin{figure*}[t]
    \begin{center}
    \includegraphics[width=1\textwidth]{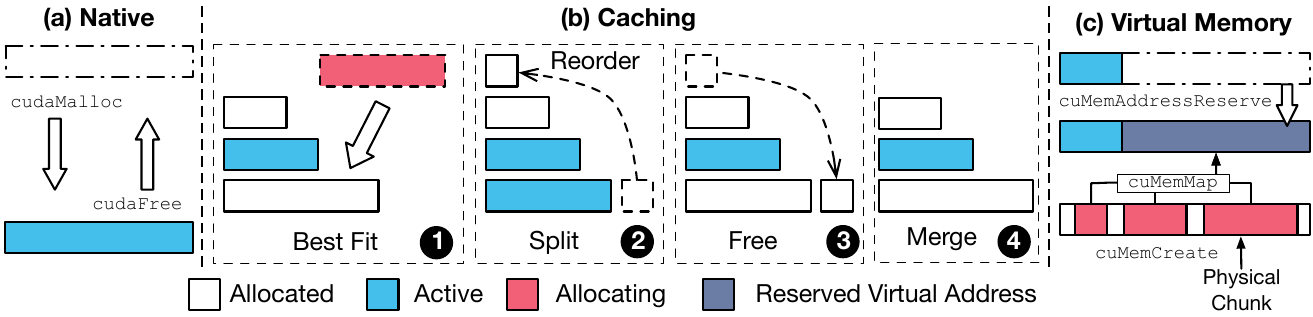}
    \caption{Three memory management strategies.}
    \label{fig:moti}
    \end{center}
\end{figure*}

\section{Background and Motivation}\label{sec:background}
In this section, we provide essential background concerning the memory management of the DL framework and outline the motivation behind conducting this study through our experimental observations.
To begin, we provide a concise overview of the increasing trend of large-scale DNN such as LLM.
Subsequently, we conduct a comparative analysis of various memory management methods.
Following the comparison, we uncover significant fragmentation challenges that arise in the context of LLMs during distributed training and memory-efficient optimization strategies.
As a result, it is imperative for us to develop a novel and efficient memory allocator that can effectively address these challenges.

\subsection{Large-scale DNNs}

LLM, such as OpenAI's GPT series~\cite{jansen2023employing, min2021recent, zhao2023survey}, represent the success of large-scale DNNs and have led to significant advancements in various language processing tasks. 
\fixme{GPT-2 represents a significant advancement over its predecessor, offering a ten-fold increase in model size and complexity}~\cite{radford2019language}, followed by GPT-3~\cite{brown2020language} with 175 billion parameters, and ChatGPT~\cite{liu2023summary}, fine-tuned for conversations.

However, the size and complexity of these models pose considerable challenges in training and deployment. 
The requirements for vast computational resources, enormous data, and extensive time (e.g., OPT-175B~\cite{zhang2022opt} taking 34 days with 1024 A100 GPU) have intensified focus on efficient training optimization. 
\fixme{Therefore, this study emphasizes the importance of efficient memory management for LLM training.}

\subsection{Memory Management of DL Framework}
\label{subsec:mm}

\fixme{Frameworks like PyTorch~\cite{paszke2019pytorch} and TensorFlow~\cite{abadi2016tensorflow} play a crucial role in DNN model training and inference. Concurrently, GPU~\cite{khalilov2021performance, choquette2020nvidia, otterness2020amd} has become an important hardware for high-performance model execution.}
This study focuses on the memory management optimization for those popular frameworks on GPUs. We compare the three types of memory management: GPU native allocator, caching allocator, and virtual memory (VM) allocator.
We conduct multiple experiments to show the efficiency associated overhead for each allocator.

\paragraph{Native Allocator.}
\label{sec:native}
As depicted in \Fig{fig:moti}(a), the native allocator is provided through GPU-vendor-supplied APIs, i.e., \texttt{cudaMalloc} and \texttt{cudaFree}, which need device synchronizations. 
The native allocator has a simplistic design that lacks flexibility. 
This makes it unsuitable for applications that need dynamic and resizable memory structures or complex memory management, especially in the context of DL.
If the DL framework implements the native GPU allocator without proper synchronization optimization, it may cause unacceptable overhead for training the DNN models.

Our experimental results have quantified the overhead of the \fixme{native} allocator.
We disable the PyTorch caching allocator (presented in the next paragraph) to train the OPT-1.3B model~\cite{zhang2022opt} on four A100-80G GPUs.
\fixme{The native allocator in PyTorch provide identical programming model for users, who can change the environment variables to set the allocator.}
The throughput of the GPU native allocator is $9.7\times$ lower than the original PyTorch allocator.
Therefore, an efficient memory management design should be one of the most critical components of the DL framework.

\paragraph{Caching Allocator.}\label{subsec:mm_ca}
DL frameworks usually use the caching allocator with a memory pool for fast memory allocation and deallocation without device synchronizations.
\Fig{fig:moti}(b) depicts the BFC algorithm~\cite{tensorflow2022} in the caching allocator employed by PyTorch and TensorFlow. 
The BFC implementations of PyTorch and TensorFlow are almost the same, with minor differences in their data structures. 

There are four main operations in the BFC algorithm.
\circlednumber{1} It begins with searching for the most suitable allocated but inactive memory block, known as the ``best fit''.
If there is no suitable allocated memory block candidate, the caching allocator invokes native GPU allocator APIs to allocate new memory blocks.
\circlednumber{2}
If the requested memory is smaller than the best-fit block, the algorithm splits the block into two blocks to boost memory utilization.
\fixme{One of the split blocks is allocated to fulfill the memory request, while the other remains in the memory pool for future reallocation. 
To effectively manage the memory, these two blocks are interconnected through a bidirectional link, with each block monitoring the availability status of its adjacent block.}
\circlednumber{3} For the free (deallocation) operation, the algorithm does not invoke the native GPU API \texttt{cuMemFree} but only releases the assignment (pointer) to the block and sets the block to inactive.
\circlednumber{4} 
Finally, the caching allocator examines whether the blocks adjacent to the left or right are also inactive. If so, the caching allocator would merge those adjacent inactive blocks into a single block.

Obviously, the caching allocator can significantly reduce the invocations of native GPU memory allocator APIs.
In the best case, all memory blocks are allocated and deallocated only once through the native allocator.
As such, the caching allocator can be much more efficient than the native GPU allocator and \fixme{is} widely adopted in existing DL frameworks.
On the other hand, the ``splitting'' mechanism will provoke a lot of possible memory fragmentation issues when the memory allocation requests are irregular and their sizes are significantly different from each other.
This fragmentation issue is not notable previously because the models are usually regular and not large enough.
For example, Transformer-based model~\cite{vaswani2023attention} is a stack of multiple identical layers with the same size of tensor, leading to minor memory fragmentation.

However, as the size of LLM grows, the fragmentation issue significantly deteriorates due to the distributed training and complex training memory strategy, leading to limited batching and inefficient memory management and training.
In the next subsections, we observe fragmentation issues become challenging in these complex training scenarios.

\begin{figure}[b]
    \begin{center}
        \minipage{0.48\linewidth}
           \includegraphics[width=\linewidth]{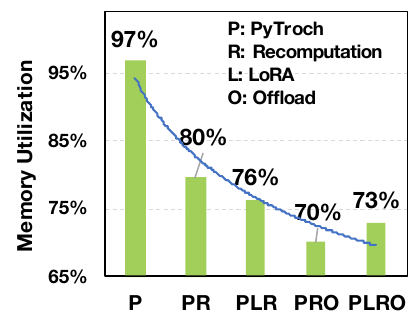}
        \caption{Memory utilization with five method combinations.}
        \label{fig:PLRO}
    \endminipage\hfill
    \minipage{0.48\linewidth}
     \includegraphics[width=\linewidth]{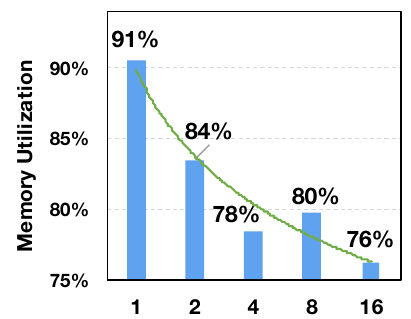}
    \caption{Memory utilization with different GPU numbers.}
    \label{fig:mul-GPU}
    \endminipage
\end{center}
\end{figure}

\subsection{Memory-efficient Optimization}\label{sec:memory_optimization}
The rapid growth in the memory requirements of large-scale DNN models~\cite{sun2019gradientflow, ahn2022ai} has sparked the development of methods at the system- and algorithm-level to alleviate memory demands. 
Examples for these methods include recomputation~\cite{kusumoto2019graph}, offloading~\cite{ren2021zero}, and low-rank adaptation~\cite{hu2021lora}.

Recomputation~\cite{kusumoto2019graph, yang2019recomputation}, also known as checkpointing, involves recalculating specific layer outputs during backpropagation rather than storing them, allowing for memory savings. 
Offloading (also known as swap), such as ZeRO-Offload~\cite{ren2021zero} shifts optimizer memory and computation from GPU to CPU, enabling training of large models on a single GPU and scaling across multiple GPUs.
In addition to these system-level approaches, algorithmic methods can also effectively reduce memory requirements.
For example, LoRA~\cite{hu2021lora}, designed for large-scale models, introduces rank-decomposition matrices to minimize both trainable parameters and GPU memory requirements, achieving accuracy similar to full model fine-tuning with reduced cost and time.

However, even though these optimizations can effectively reduce memory footprint in GPU memory, they may sometimes lead to poor memory utilization.
As depicted in \Fig{fig:PLRO}, we train the OPT-1.3B model on four A100-80G GPUs with different optimization method combinations. 
Using only PyTorch (P) achieves high memory utilization, whereas employing techniques such as LoRA (L), Recomputation (R), or Offload (O) significantly reduces memory utilization.
According to our investigations, combining these techniques results in high memory fragmentation.
The reason is that these memory optimization techniques inherently incur dynamic and irregular allocation requests.

\begin{figure}[b]
    \begin{center}
    \includegraphics[width=0.49\textwidth]{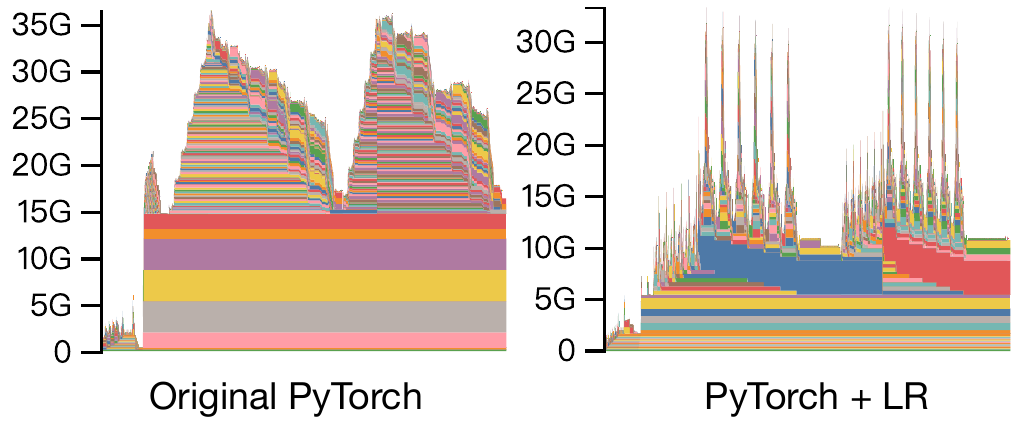}
    \caption{\fixme{Memory footprint of GPT-Neox-20B training.}}
    \label{fig:irregularity}
    \end{center}
    \vspace{1mm}
\end{figure}

\fixme{
To explore the origin of this irregularity, we present the memory footprint of a training process on the GPT-NeoX-20B model.
As shown in \Fig{fig:irregularity}, the left figure shows the footprint on the original PyTorch, and the right figure is collected from PyTorch with LR (LoRA and Recomputation) optimization.
Obviously, the right figure shows more irregularity than the left because of the usage strategy like recomputation. 
Statistically, the left figure makes 46 thousand allocations with a size of 93 MB on average, while the right figure has 76 thousand allocations with 85 MB on average, indicating that complex strategies lead to more frequent and smaller allocations thus causing fragmentation.
}

The results motivate us to address the memory fragmentation issues for more efficient and scalable large-scale DNN model training.

\noindent\textit{\textbf{Observation 1}: The more complex and irregular memory optimization strategies used, the more fragmentation there will be.}

\subsection{Distributed Training}\label{sec:distributed_training}

With the increasing complexity of DNN models, distributed training has become vital, especially for LLMs. 
Data parallelism, supported by PyTorch distributed data parallel (DDP) \cite{li2020pytorch}, duplicates the setup to process different portions of data simultaneously, synchronizing after each training step. 
Model parallelism splits the model across multiple GPUs, each handling different stages. 
The model parallelism includes two categories: pipeline parallelism \cite{lepikhin2020gshard, shoeybi2019megatron, xu2023efficient}, placing individual layers on single GPUs, and tensor parallelism~\cite{harlap2018pipedream, huang2019gpipe, li2021chimera}, dividing each tensor into chunks for specific GPUs.

Obviously, using more GPUs would cause more fragmentation due to its irregular memory allocation and deallocation. 
To examine this effect, we conduct a test implemented on OPT-13B on PyTorch.
As depicted in \Fig{fig:mul-GPU}, when the number of GPUs is one, the memory utilization is {>90\%}. 
However, as the number of GPUs grows to 16, the memory utilization declines to {76\%}, even though multi-GPU parallelism is essential for training large models. 
Such fragmentation wastes memory resources and limits the batch size of LLM training.

\noindent\textit{\textbf{Observation 2}: As the number of GPUs scales up, the issue of memory fragmentation is likely to become more pronounced. }

\begin{figure}[t]
    \begin{center}
    \includegraphics[width=0.48\textwidth]{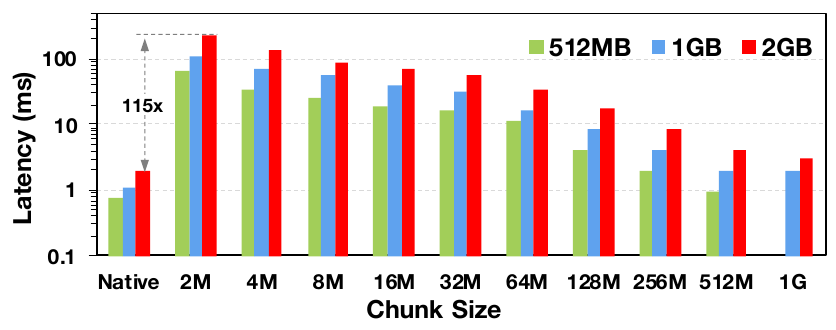}
        \caption{The allocation latency of native allocator (the first one) and virtual memory allocator.}
        \label{fig:virtual}
    \end{center}
\end{figure}

\subsection{Low-level Virtual Memory (VM) Management}
\label{subsec:vmm}

Recognizing the growing need among applications to manage memory quickly and efficiently, CUDA has introduced a new feature called low-level virtual memory management~\cite{perry2020introducing}, similar to \fixme{Windows's} VirtualAlloc~\cite{virtualalloc2022} and Linux's mmap~\cite{mmap2}. 
This feature breaks the malloc-like abstractions, and offers primitive operations such as reserve and map to manipulate the virtual address space. 
In our work, we show that this low-level VM management can be used to reduce memory fragmentation and improve memory utilization for large-scale DNN training, which we call as virtual memory allocator.ß

\begin{table}[t]
    \vspace{2mm}
        \centering
            \resizebox{0.8\linewidth}{!}{
            \begin{tabular}{l|c|c|c}
            \toprule
            {Chunk Size} & 2~MB & 128~MB & 1024~MB\\
            \midrule
            \texttt{cuMemReserve} & 0.003 & 0.003 & 0.002\\
            \midrule
            \texttt{cuMemCreate} & 18.1 & 0.89 & 0.79\\
            \midrule
            \texttt{cuMemMap} & 0.70 & 0.01 & 0.002\\
            \midrule
            \texttt{cuMemSetAccess} & 96.8 & 8.2 & 0.7\\
            \midrule
            {Total} & 115.4 & 9.1 & 1.5\\
            \bottomrule
            \end{tabular}
        }
        \vspace{2mm}
        \caption{The VMM API execution time breakdown normalized by \texttt{cuMalloc} execution time.}
        \label{tbl:vmmBreakdownTable}
    \end{table}
    
\Fig{fig:moti}(c) illustrates the basic idea of using this low-level virtual memory management.
The \texttt{cuMemAddressReserve} function reserves a virtual memory address for the new memory allocation, and \texttt{cuMemCreate} allocates physical memory chunks on GPU.
\fixme{It is not revealed how the underlying system translates memory in the physical address space. Furthermore, there is no guarantee of contiguity of physical chunks.}
The \texttt{cuMemMap} function bridges the physical and virtual memory, mapping the physical handle to the reserved address.
CUDA also offers a suite of memory deallocation functions such as \texttt{cuMemUnmap}, \texttt{cuMemAddressFree}, and \texttt{cuMemRelease}. 
Obviously, the advantage of low-level VM API is that it can allocate and map the non-contiguous physical chunks, which can tackle GPU memory fragmentation issues.
However, the virtual memory allocator costs much more expensive overhead than native GPU allocator.

To verify the overhead of the VM allocator, we have the experiments on memory allocation with three different sizes: 512~MB, 1~GB, and 2~GB, which are total allocated block sizes.
\Fig{fig:virtual} illustrates the comparative results between the native memory allocator and the virtual memory allocator. 
The y-axis represents the allocation latency, taken on a logarithmic scale. 
On the x-axis, 2~MB, 4~MB, \dots, and 1024~MB represent the sizes of internal physical chunks that construct the allocation block.
For example, the 1~GB allocation block needs to map 512 chunks with 2~MB size.
Finally, the result in \Fig{fig:virtual} shows that the latency of virtual memory is exceedingly high. 
Specifically, if the virtual memory block is partitioned into 2~MB chunks, it would be over $100\times$ slower than the native allocator, which is totally unacceptable.

\fixme{To further explore the bottleneck of the VMM API, we provide the execution time breakdown of allocation. 
\Tbl{tbl:vmmBreakdownTable} shows the latency breakdown of the VMM API with 2~GB GPU memory allocation. All latency is normalized to the \texttt{cuMalloc}. 
Each allocation in GMLake needs only one \texttt{cuMemAddressReserve} but multiple \texttt{cuMemCreate}, \texttt{cuMemMap}, and \texttt{cuMemSetAccess} for each physical chunk. \texttt{cuMemSetAccess} is a special function to make the map available provided by VMM. We can see that using a 2MB small chunk to allocate 2 GB of memory is $115\times$ slower than native \texttt{cuMalloc}.
}

\noindent\textit{\textbf{Observation 3}: Although virtual memory can reduce memory fragmentation, the original virtual memory allocator on GPU still presents many challenges and \fixme{needs} further optimization.}

\section{\proj{}}

\begin{figure}[t]
    \begin{center}
    \includegraphics[width=0.48\textwidth]{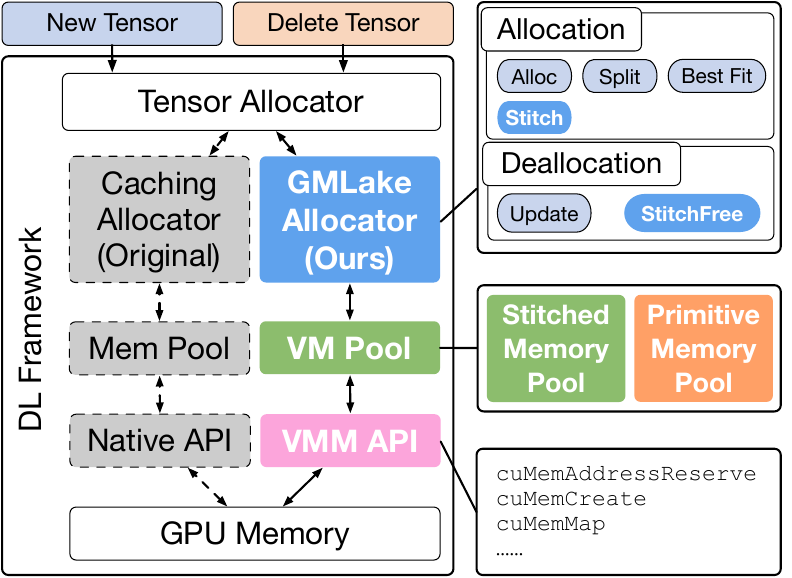}
    \caption{The overview of caching allocator and \proj{}.}
    \label{fig:overview}
    \end{center}
\end{figure}

In this work, we introduce \proj{}, an efficient memory allocation method specifically designed for GPU memory (i.e., GPU memory lake). \proj{} leverages CUDA's low-level VM management features to expedite memory allocation and deallocation processes.
\Fig{fig:overview} provides an overview of \proj{}, which provides the same memory (de)allocation interfaces to the existing caching allocator but internally integrates the virtual memory stitching (VMS) mechanism. 
This integration is achieved through the precise utilization of CUDA's low-level VM management APIs.

The original caching allocator in DL frameworks adopt the BFC algorithm, as explained in \Sec{subsec:mm_ca} and shown in \Fig{fig:moti}(b).
To avoid synchronization and improve memory management efficiency, those frameworks internally manage the memory pool to handle the (de)allocation instead of directly using the native APIs.
Following this approach, our \proj{} also has the three following components: 

\textbf{Virtual memory API}: This refers to the low-level APIs employed to instruct the GPU to allocate and free memory using virtual memory addresses, a process characterized by significant overhead if not fully optimized.

\textbf{Virtual memory pool}: Serving as the foundational data structure, this is designed for caching virtual memory. Its implementation is crucial for enhancing efficiency.

\textbf{\proj{} allocator}: This includes all functions, algorithms, and strategies essential for managing the VM pool.

In this section, we describe the design and details for the three integral components and assemble the \proj{} framework, proceeding from the foundational to the topmost layer.

\subsection{Virtual Memory API}
\label{subsec:api}

As introduced in \Sec{subsec:vmm} and illustrated in \Fig{fig:moti}(c), the low-level VMM APIs serve as the fundamental interface between the GPU and the application. 
As depicted in the bottom of \Fig{fig:vmp}, we utilize the VMM API to build the primitive block (pBlock), a critical data structure for the \proj{} allocator. 
The operations within the pBlock include:

\textbf{\texttt{AddrReserve}}: Initially, the allocation of a primitive block necessitates specifying the allocation size and reserving the corresponding virtual address (VA).

\textbf{\texttt{Create}}: Following that, the primitive block creates the physical chunks where the data is stored physically.

\textbf{\texttt{Map}}: Lastly, the primitive block maps all physical chunks to the virtual address, enabling seamless access for the tensors.

To optimize defragmentation, we apply a uniform chunk size of 2~MB across all chunks. 
While the overhead of this 2~MB chunk size is considerable (\Sec{subsec:vmm}), it can be mitigated through an efficient data structure and a well-designed stitching strategy \fixme{that} 
achieves the best defragmentation effect and the best compatibility with PyTorch code base. Meanwhile, we use the following DNN-specific optimization to reduce the frequency of stitching so that its end-to-end overhead is negligible.
\fixme{\proj{} uses VMM to tackle allocation larger than 2MB. For memory allocation less than 2MB, we use the original PyTorch splitting method of the caching allocator to deal with its internal fragmentation issues. Moreover, allocation < 2MB is rare in LLM training.}

The map operation serves as the basic operation for virtual memory stitching, allowing us to concatenate multiple physically continuous chunks that may not be contiguous in physical memory. 
Employing the VMM APIs, we can orchestrate the virtual memory into the virtual memory pool, the underlying data structure for the \proj{} allocator.
The details of VM pool are presented as follows.

\subsection{Virtual Memory Pool}

\begin{figure}[t]
    \begin{center}
    \includegraphics[width=0.48\textwidth]{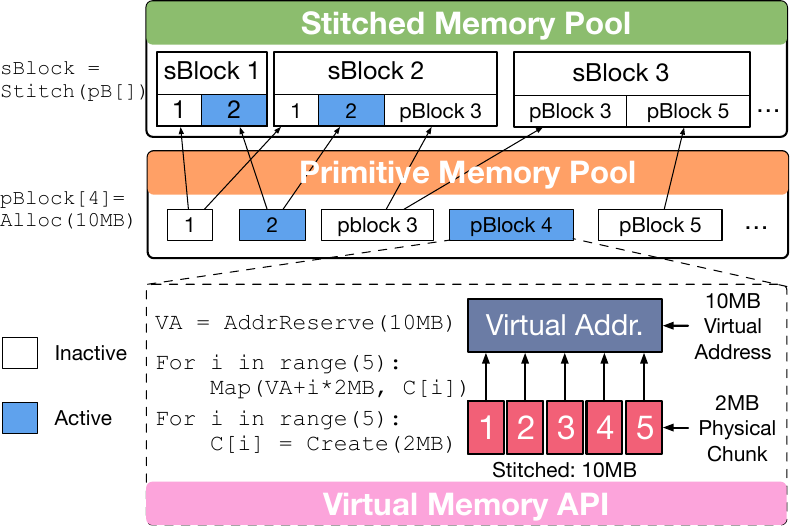}
    \caption{The data structure of primitive and stitched memory pool.}
    \label{fig:vmp}
    \end{center}
\end{figure}

Given that original VMM APIs are time-consuming, it \fixme{is} crucial to reduce their usages to achieve high efficiency for \proj{}. 
Drawing inspiration from the caching allocators, we have designed our virtual memory pool (VMP) with caching capability, thereby markedly reducing the instances of physical memory (de)allocations. 
Shown in \Fig{fig:vmp}, we differentiate between two types of memory pools: the primitive memory pool (pPool) and the stitched memory pool (sPool).

\paragraph{pPool and pBlock.}
The data structure of pPool utilizes a sorted set to store the pBlocks. 
For each pBlock, pPool begins by constructing a structure to document the pointer to the pBlock, including essential basic attributes, such as the active state of the pBlock. 
Subsequently, the newly allocated pBlock is inserted into the set, with all pBlocks sorted by block size in descending order. 
The pBlock, serving as the primitive block, represents the smallest unit accessible to high-level tensors, and it functions as a basic data structure that can be stitched and pointed to by multiple sBlocks.

\paragraph{sPool and sBlock.}
The sPool is also organized into a sorted set, similar to pPool.
Its elements comprise the stitched block structure, which integrates multiple pBlocks. 
For instance, as illustrated in \Fig{fig:vmp}, sBlock 3 contains pBlocks 3 and 5, which are stitched together, and pBlock 3 may also be pointed to and stitched by sBlock 2. 
Consequently, the attributes of the sBlock are influenced by the pBlocks, such that if even one pBlock is active, all corresponding sBlocks are labeled as active. 
In practice, the sBlock remaps virtual memory to all physical chunks of the pointed pBlocks, making it accessible by high-level tensors. 
To streamline the process, we stipulate that sBlock can only be allocated or assigned to tensor allocations that match the sBlock size. 
Further details on this constraint are provided in the subsequent section.

Comparing the physical chunk, pBlock, and sBlock reveals their roles as data structures at different levels. 
While physical chunk is controlled by the low-level API and remains transparent to the high-level tensors, pBlock and sBlock reside in the pPool and sPool, respectively, offering the virtual memory address for access by high-level tensors. 
Moreover, sBlock operates at a more advanced level, consisting of multiple pBlocks. 
Next, we describe how \proj{} uses those data structures to achieve efficient memory management.

\subsection{Allocator}
The allocator includes all the essential functions and algorithms for the memory allocation and deallocation. 
Due to the space limit, we only briefly explain the most important functions used in the allocation and deallocation modules.

\subsubsection{Allocation Module}
\label{sec:allocationModule}

The \textbf{\texttt{Alloc}} function is responsible for allocating a new pBlock and inserting it into pPool, as detailed in \Fig{fig:vmp}. It serves as the exclusive interface for allocating new physical chunks and incrementing the allocated GPU memory.

The \textbf{\texttt{Split}} function divides a pBlock (primitive block) into two smaller pBlocks, similar to the ``Split'' operation depicted in \Fig{fig:moti}, but with an entirely distinct underlying implementation. Specifically, the \texttt{Split} function in \proj{} operates based on the pBlock structure, resulting in two new pBlocks with corresponding virtual memory addresses and remapped physical chunks. The previous pBlock structure is subsequently removed from the pPool set.

The \textbf{\texttt{Stitch}} function is the sole mechanism to create an sBlock and insert it into the sPool, as shown in \Fig{fig:vmp} top. This function is an integral component of our allocator and can stitch together multiple pBlocks into a single sBlock. 
\fixme{
We use the VMM API, i.e., the low-level API provided by NVIDIA specifically for Virtual Memory Management (VMM), to ``stitch'' two pBlocks, as shown in \Fig{fig:moti} (c). 
Assume we have two pBlocks, $p_1$ (1GB) and $p_2$ (2GB). We adopt the VMM API \texttt{cuMemCreate} to create corresponding physical chunks and reserve virtual address (VA) by \texttt{cuMemAddressReserve}. 
Then, VA maps PA using \texttt{cuMemMap}. Actually, we do not need to unmap the original VA-PA mapping for $p_1$ and $p_2$, as the PA in VMM can be pointed by multiple VAs. Therefore, we only reserve a 3GB VA of the sBlock $s_1$ using \texttt{cuMemAddressReserve}. 
For all sBlocks, they never create \texttt{cuMemCreate} new  physical chunks. We use the API \texttt{cuMemMap} to map the VA of $s_1$ (3GB) to the physical chunks of $p_1$ and $p_2$. Since multiple sBlocks can contain the same physical chunks, we need more attributes to ensure that each physical chunk is used by only a single tensor, such as the active state of the pBlock.
}

\begin{algorithm2e}[t]
    \small
       \DontPrintSemicolon
       \KwIn{
           Allocating block size: $bSize$;\\
           \ \ \ \ \ \ \ \ \ \ \ \ \ Inactive sBlocks and pBlocks: ${sBlocks, pBlocks}$.
       }
       \KwOut{
           State, $state$; Candidate pBlocks, $CB$.
       }
   
       \SetKwFunction{FMain}{BestFit}
       \SetKwProg{Fn}{def}{:}{}
       \Fn{\FMain{$bSize$, $sBlocks$, $pBlocks$}}{
            \tcp*[h]{SBlock only for S1: Exact match.}\\ 
            \ForEach{$block \in sBlocks \cup pBlocks$}
            {
                \If{block.size == bSize}
                {
                    \Return{1, [block]}
                }
            }
            \fixme{CB=[]}\\
            $CBSize$ = 0\\
            \ForEach{$block \in pBlocks$}
            {   
                \If{block.size $\geq$ bSize}
                {
                    $CB = [block]$\\ $CBSize = block.size$
                }
                \ElseIf{CBSize < bSize}
                {
                    $CB$.append($block$)\\ $CBSize$ += $block.size$
                }
               \fixme{ \Else
                {
                    break
                }
               }
            }

            \If{CB.length == 1 \& CBSize > bSize}
            {
                \Return{2, CB}                 \tcp*[h]{S2: Single block.}\\ 
            }
            \ElseIf{CBSize $\geq$ bSize}
            {                
                \Return{3, CB} \tcp*[h]{S3: Multiple blocks.}
            } 
            \Else
            {
                \Return{4, CB}  \tcp*[h]{S4: Insufficient blocks.}
            }
       }
       \caption{The \texttt{BestFit} Function.}
       \label{alg:bestFit}
\end{algorithm2e}

The \textbf{\texttt{BestFit}} function identify the most suitable pBlock or sBlock for memory allocation, returning the state and candidate blocks for subsequent processing.
As detailed in Algorithm~\ref{alg:bestFit}, we have designed four states, covering all scenarios \proj{} may face. 
It operates on the assumption that both sPool and pPool are sorted in descending order of size.

\begin{itemize}
    \item \textbf{Exact match} (Line 2-4): This state arises when the size of a candidate block matches the allocation size. The block may be either an sBlock from sPool or pBlock from pPool. This is the sole situation where an sBlock can be assigned for new allocation. All other states exclusively involve pBlock.

\item \textbf{Single block} (Line 12): Here, \texttt{BestFit} identifies the best-fit (minimum) pBlock that is larger than the requested allocation size.

\item \textbf{Multiple blocks} (Line 14): In scenarios where all pBlocks are smaller than the required allocation size while their total size satisifies allocation requirment, the \texttt{BestFit} function greedily seeks multiple candidate pBlocks to stitch together.

\item \textbf{Insufficient blocks} (Line 16): This state occurs when there is no enough pBlocks to meet the requested allocation size, though \texttt{BestFit} still returns a block list.
    
\end{itemize}

\fixme{
\proj{} substitutes several of the internal functions of the PyTorch caching allocator module. The ``stitch'' operation is completely user-transparent and doesn't burden user to modify their code.
The CUDA VMM API refers to the low-level API provided by NVIDIA specifically for Virtual Memory Management (VMM).The CUDA API we adpot not only includes the VMM-related, but also comprises regular ones like cuMemAlloc. The implementation of \proj{} APIs leverages CUDA APIs to achieve fine-grained memory stitching and reusing, thus they belong to distinct levels and serve corresponding functionalities.
}

\subsubsection{Deallocation Module}

The deallocation module refrains from actively deallocating physical GPU memory using the low-level VMM API. Instead, it only updates or restores the stitched virtual memory blocks.

\paragraph{\texttt{Update}.}
Upon receiving a deallocation request for a high-level tensor, we substitute the original VMM deallocation function with the \texttt{Update} function. 
This function alternates the state of active pBlocks and sBlocks, thereby effecting the removal of links and assignments between tensors and blocks. 
Throughout the program runtime, actual physical memory remains under the control of corresponding pBlock.

\paragraph{\texttt{StitchFree}.} 
The purpose of this function is to release the Least Recently Used (LRU) sBlocks held within the sPool. 
Due to space limitations, we skip intricate details here. 
We have implemented complete algorithms and data structures to support the LRU-based \texttt{StitchFree}.
Notably, we only release the inactive sBlock structure from sPool.

We have presented details for various components in \proj{}, including low-level APIs, data structures, and high-level functions.
Leveraging these thoughtfully engineered features, we are equipped to execute efficient allocation strategies within \proj{}, which effectively tackle memory defragmentation challenges in large-scale DNN training.

\section{Defragmentation Strategy}
In this section, we present the strategies to reduce the memory fragmentation issue.
We first propose a sophisticated algorithm that theoretically eliminates all fragmentation based on \proj{} allocator.
We then discuss and describe our optimizations to guarantee its efficiency and robustness.

\begin{figure}[t]
    \begin{center}
    \includegraphics[width=0.48\textwidth]{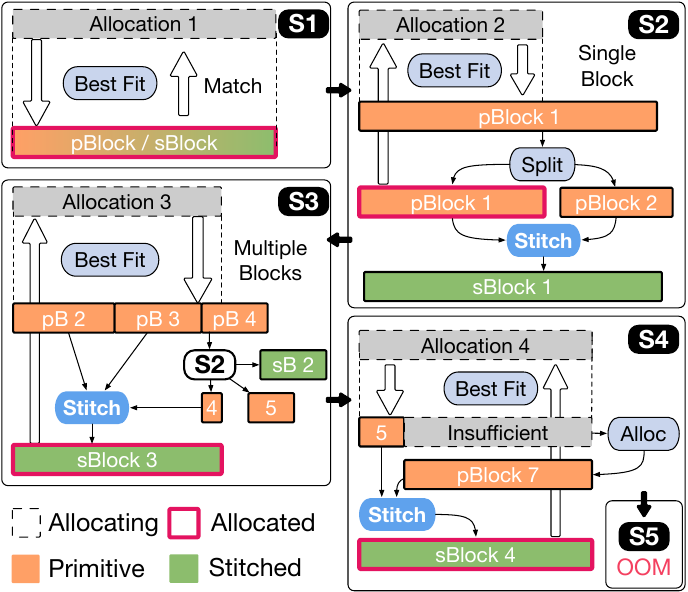}
    \caption{Memory allocation  strategy in \proj{}.}
    \label{fig:strategies}
    \end{center}
\end{figure}

\subsection{Memory Allocation Strategy}
\Fig{fig:strategies} shows the \proj{} allocation strategy to allocate or assign a new memory block to a new tensor allocation request.
This strategy is based on the four states provided by \texttt{BestFit} module, for each of which we design  a post-processing step.

In state \circlednumber{S1}, an immediate return of existing pBlock or sBlock is made for Allocation 1. 
If an exact matching block is not found within the inactive sPool and pPool, we progress to state \circlednumber{S2}, guided by a single block produced by the \texttt{BestFit} function. 
This requires the partitioning (via \texttt{Split}) of the larger pBlock 1 into two distinct pBlocks, both inserted into the pPool. 
The newly created pBlock 1 replaces its predecessor and is allocated for Allocation 2. 
Simultaneously, \proj{} executes a \texttt{Stitch} function, merging pBlock 1 and pBlock 2 into sBlock 1, which is then added to the sPool.

In state \circlednumber{S3}, \proj{} engages in merging (\texttt{Stitch}) multiple pBlocks to create consolidated sBlock 3 for Allocation 3. 
If needed, the final pBlock can be subdivided (\texttt{Split}) similar to the procedure in \circlednumber{S2}. 
The stage concludes with the introduction of new pBlocks, specifically pBlock 4 and pBlock 5, into pPool, while sBlock 2 and sBlock 3 are added to sPool.

In state \circlednumber{S4}, it is possible that available blocks for stitching and allocation are insufficient for Allocation 4. 
\proj{} then triggers the \texttt{Alloc} function, using the low-level API to create pBlock 7 with new physical chunks and corresponding virtual memory addresses. This new pBlock is added to the pPool. 
Additionally, we merge (via \texttt{Stitch}) pBlock 5 and pBlock 7 into a new sBlock 4, which is returned for Allocation 4 and added to the sPool. 
If no eligible candidate blocks are present (i.e., the absence of pBlock 5), \proj{} directly allocates pBlock 7, without using the \texttt{Stitch} function. 
If the \texttt{Alloc} function call fails, \proj{} immediately reports an Out-of-Memory (OOM) error in state \circlednumber{S5}.

\subsection{Strategy Analysis}
We analyze the \proj{} memory allocation strategy from aspects of effectiveness, efficiency, and robustness.

\subsubsection{Effectiveness}

The \proj{} allocation strategy effectively ensures that nearly all fragmentation is eliminated from the GPU system.

\paragraph{Interface.} 
The effectiveness of this strategy is aided by our interface design, which consolidates all operations into three main functions: \texttt{Alloc}, \texttt{Split}, and \texttt{Stitch}. 
\texttt{Alloc} is the only function that can create a new pBlock, and only \texttt{Stitch} can generate a new sBlock, while \texttt{Split} does not increase the allocated memory.

\paragraph{Data Structure.} 
The pPool represents a strict one-to-one mapping of GPU memory, with each pBlock being distinct from others. 
It is an allocated GPU memory set devoid of duplicate elements and overlapping addresses. 
The sPool is designed to store sBlocks, reserving links and pointers to the pBlocks in a manner similar to a soft link mechanism. 
\proj{} prohibits the splitting of sBlocks as it may affect the pBlocks. The sPool is considered a subset of the pPool.

In the end, each time the program reaches a new peak in GPU memory usage, for example, when calling the \texttt{Alloc} function, the pPool may not provide enough blocks for stitching and allocation, resulting in full memory utilization without fragmentation. 
This contrasts with the original caching allocator, which may leave many sub-blocks unused.

\subsubsection{Efficiency}

We incorporates several methods to achieve high efficiency. 
Initially, the algorithmic problem in \circlednumber{S3} represents a classic NP-hard packing problem~\cite{martello2003exact}. 
Yet, through the application of the \texttt{Split} and \texttt{Stitch} functions, an exactly-matched block is generated to fit the allocation, resulting in linear complexity.

Secondly, the union set comprising sPool and pPool chronicles all sizes and corresponding blocks for every tensor allocation, akin to a tape recording the tensor allocation pattern for DNN models. 
Fortunately, DNN model training is highly regularized, as each iteration processes identical model parameters and input data sizes. 
Therefore, after a few iterations, \proj{} will no longer execute \circlednumber{S2}, \circlednumber{S3}, and \circlednumber{S4}. 
\proj{} will only utilize the \circlednumber{S1} ``exact match'' strategy for the remainder of the training, contrasting with the original caching allocator, which continuously requires splitting and merging operations.

\subsubsection{Robustness}

In practice, stitching and creating new sBlocks cannot occur infinitely due to the total capacity limitation on the GPU. 
Moreover, excessive stitching operations can impair the efficiency of the \proj{} allocator when running allocation modules on sPool, such as \texttt{BestFit}. 
When the total capacity surpasses this limitation or threshold, \proj{} employs \texttt{StitchFree} to release the LRU sBlocks and clear the pattern tape, thereby serving as a fallback mechanism for robustness.

Furthermore, when DNN training exhibits an extremely irregular pattern, it may generate numerous small blocks leading to frequent splits and stitches, causing early attainment of the limitation. 
To avoid unnecessary performance loss, a minimal fragmentation limit is established. 
If a block is smaller than this limit, \proj{} will avoid stitching or splitting it. 
Hence, all algorithms and modules adhere to the fragmentation limit (e.g., 128~MB), to ensure high efficiency and robustness in DNN training.

\section{Evaluation}
\label{sec:eval}

We implement \proj{} with 5000 lines of C++ code and integrate it into the  caching allocator of PyTorch. We have adapted \proj{} to different versions of PyTorch, such as PyTorch-1.13.1 and PyTorch-2.0.

We evaluate the performance of \proj{} on fine-tuning several popular LLMs.
We compare \proj{} virtual memory allocator and PyTorch's caching allocator under diverse conditions, considering distinct training frameworks, GPU scalability, and combinations of optimization strategies.
Through this analysis, we demonstrate the scalability of \proj{} to adapt seamlessly to complex environments and its effectiveness in resolving memory fragmentation issues. 
Collectively, \proj{} achieves a significant reduction in the fragmentation ratio of \fixme{15\% on average} and  up to \fixme{$\mathbf{33\%}$}, as well as a decrease in reserved GPU memory of \fixme{9.2GB on average and} up to \fixme{\textbf{25GB}, which is obtained from 76 workloads within 8 different models}. 

\subsection{Evaluation Methodology}

\paragraph{Testbed.} 
The evaluation of \proj{} is conducted utilizing two distinct setups: single-node multi-GPU and multi-node multi-GPU experiments.
The single-node evaluations are performed on a server equipped with an Intel Xeon Platinum 8369B CPU boasting 1~TB of DRAM and eight NVIDIA A100 GPUs (80~GB memory for each). 
These GPUs are interlinked via NVLink, running on CUDA 11.4 and cuDNN 8.5. Correspondingly, the multi-node evaluations encompass two servers, each mirroring the configuration of the single-node server. In the finetuning phase, these two nodes engage in distributed training facilitated by RDMA.

\paragraph{Training Scenarios.}
We evaluate \proj{} across multiple prominent LLM optimization platforms, including Deepspeed~\cite{rasley2020deepspeed} and FSDP~\cite{zhao2023pytorch}, encompassing diverse optimization strategies such as LORA~\cite{hu2021lora}, gradient-checkpointing(i.e. Recomputation)~\cite{salimans2017gradient}, and offload~\cite{beaumont2021efficient, ren2021zero}.
We focus on the fine-tuning scenario.

\begin{table}[t]
\centering
    \resizebox{\linewidth}{!}{
\begin{tabular}{l|c|c}
\toprule
\textbf{Model} & \textbf{Strategies} & \textbf{DDP Framework}  \\
\midrule
\textbf{OPT-1.3b~\cite{zhang2022opt}} & L~\cite{hu2021lora} R~\cite{yang2019recomputation} O~\cite{ren2021zero} & Deepspeed~\cite{DeepSpeed2023}  \\
\textbf{GPT-2~\cite{radford2019language}} & R O & Colossal-AI~\cite{li2021colossal}  \\
\textbf{GLM-10b~\cite{du2021glm}} & R O & FSDP~\cite{zhao2023pytorch}  \\
\textbf{OPT-13b}~\cite{zhang2022opt} & L R O & Deepspeed  \\
\textbf{Vicuna-13b~\cite{vicuna2023}} & L R O & Deepspeed  \\
\textbf{GPT-NeoX-20b~\cite{black2022gptneox20b}} & L R O & Deepspeed \\
\bottomrule
\end{tabular}
}
\caption{Benchmark specifications. (L: LoRA; R: Recomputation; O: Offload; FSDP: Fully Sharded Data Parallel.)}
\label{tbl:benchmark}
\end{table}

\paragraph{Models and Datasets.} 
We conduct evaluation on a set of representative open-source LLMs from their official examples.
The exhaustive list of evaluated models, datasets, distributed data parallel (DDP) frameworks, and optimization strategies employed during the finetuning stage is presented in \Tbl{tbl:benchmark}. Notably, the selection of default datasets from open-source LLM applications is based on the consideration that memory usage during finetuning remains unaffected by dataset quality.

\paragraph{Baselines.}
We compare \proj{} against PyTorch 2.0 with various LLMs training, e.g., Deepspeed~\cite{rasley2020deepspeed} and Colossal-AI~\cite{li2021colossal}. To our best knowledge, PyTorch caching allocator represents a state-of-the-art, application-agnostic memory allocator that demonstrates versatile compatibility across a range of deep learning applications. 
Notably, the transition between \proj{} and PyTorch's original allocator is notably convenient by switching certain configurations. Additionally, \proj{} is equipped with a specific set of hyper-parameters, empirically configured to achieve optimal performance outcomes through best practices.

\paragraph{Metric.} \label{sec:metric}
\fixme{Different from the fragmentation metric FMFI~\cite{gorman2006and, kwon2016coordinated} for the virtual memory page with fixed length, the blocks used in \proj{} can be split with arbitrary size. Therefore, we empirically define a specific fragmentation ratio for \proj{}, which equals $(1-\text{utilization ratio})$.}
To measure the memory fragmentation, we first calculate the memory utilization ratio, which equals peak active memory divided by peak reserved memory.
The term ``active memory'' refers to the cumulative memory occupied by all active blocks, currently allocated by high-level tensors and utilized within DNN computations. 
On the other hand, ``reserved memory'' pertains to the total memory allocation set aside by both PyTorch and \proj{}. These metrics are recorded at their respective peak values. 
\fixme{To depict the relationship before and after the application of GMLake in terms of memory reduction, we calculate the arithmetic average result of the memory reduction ratio for multiple workloads. The formula to calculate the arithmetic average is $$
    Mem Reduction Ratio = \frac{\sum  Reserved-\sum GMLakeReserved}{\sum  Reserved} $$ 
where "Reserved" and "GMLakeReserved" refer to the reserved memory of PyTorch and GMLake, respectively.}
While the reserved memory typically stabilizes as the training process runs, fluctuations in active memory are more substantial due to tensor (de)allocations during model execution. 
Given the caching mechanism, the peak active memory is pertinent for utilization or fragmentation analysis.
We also employ throughput (expressed in samples per second) to quantify the speed of DNN training.

\begin{figure*}
    \centering
    \subfloat[OPT-13B.]{   
    \label{fig:opt-13b-strategy-a}
        \includegraphics[width=0.32\linewidth]{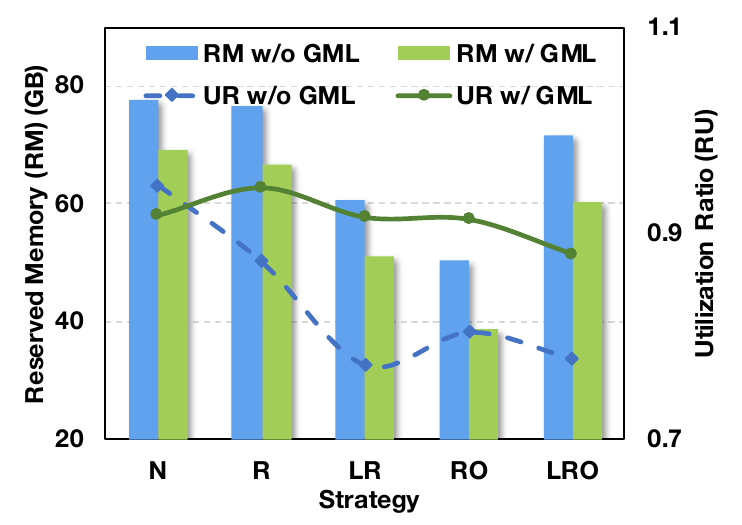}}
    \hfill
    \subfloat[Vicuna-13B.]{
    \label{fig:vicuna-13b-strategy-a}
        \includegraphics[width=0.32\linewidth]{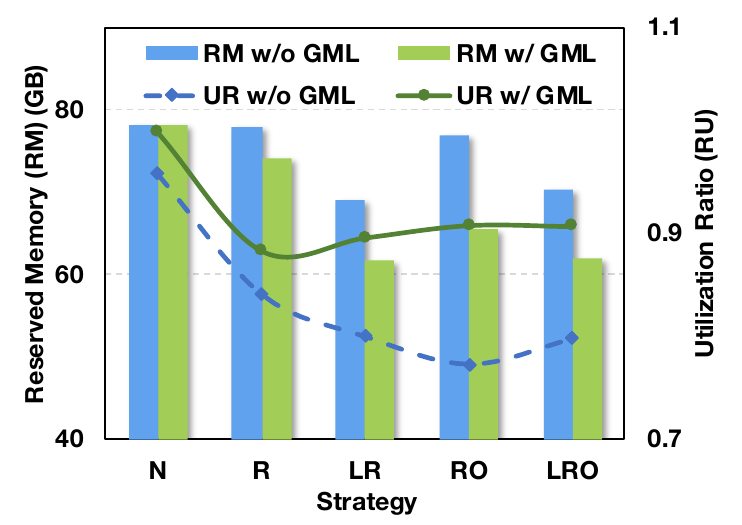}}
    \hfill
    \subfloat[GPT-NeoX-20B.]{
    \label{fig:GPT-NeoX-20b-strategy-a}
            \includegraphics[width=0.32\linewidth]{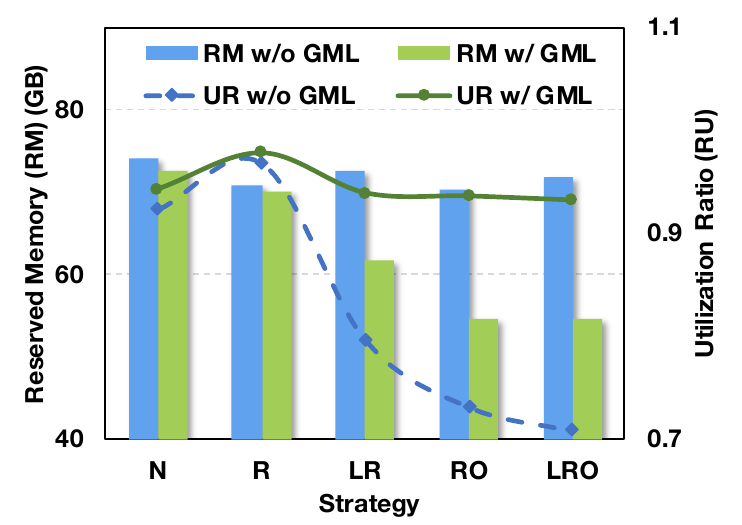}}
    \caption{Comparison of memory utilization ratio on memory-efficient strategy combinations.}
    \label{fig:eval:strategy}
\end{figure*}

\begin{figure*}[t]
    \centering
    \subfloat[OPT-13B.]{\label{fig:opt-13b-scale-up-a}
        \includegraphics[width=0.32\linewidth]{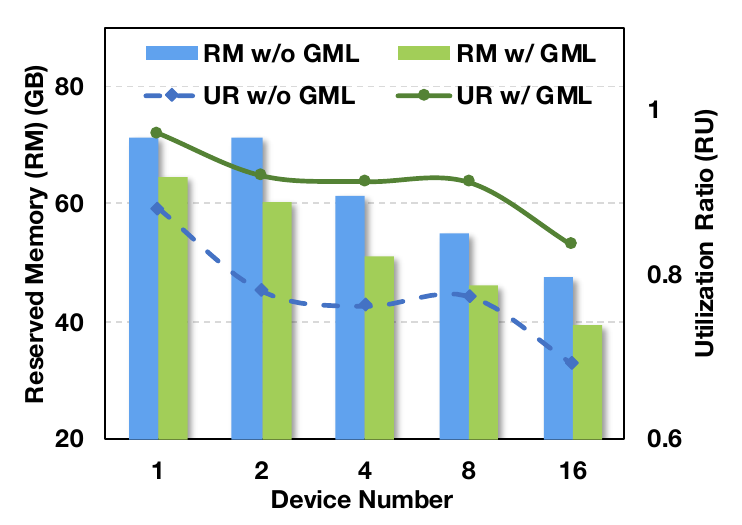}}
    \hfill
    \subfloat[Vicuna-13B.]{\label{fig:vicuna-13b-scale-up-a}
        \includegraphics[width=0.32\linewidth]{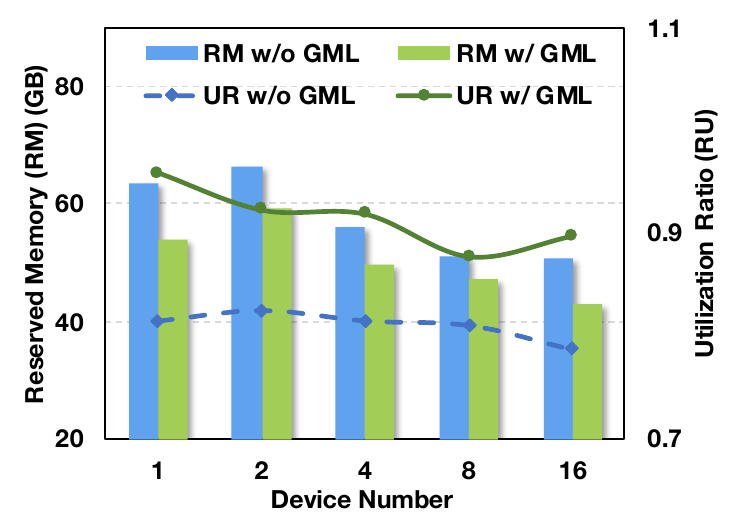}}
    \hfill
    \subfloat[GPT-NeoX-20B.]{\label{fig:eval:gpuscaleup:gpt20b}
            \includegraphics[width=0.32\linewidth]{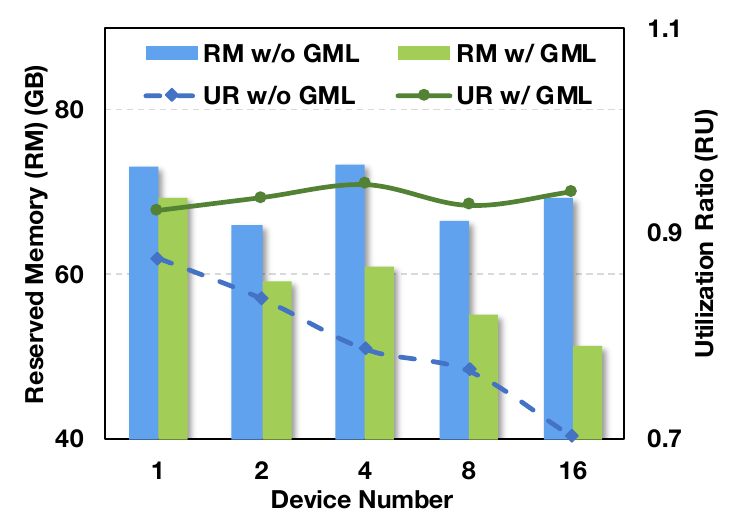}}
        \\ 
    \subfloat[OPT-13B.]{\label{fig:opt-13b-scale-up-b}
        \includegraphics[width=0.32\linewidth]{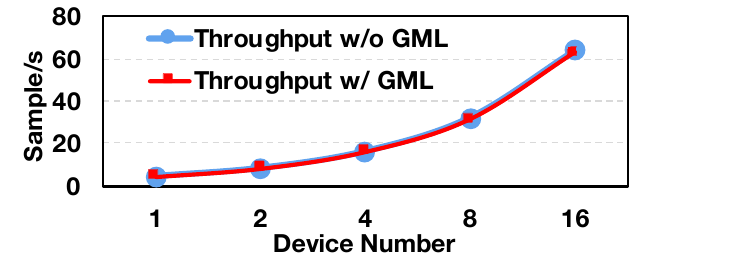}}
    \hfill
    \subfloat[Vicuna-13B.]{\label{fig:vicuna-13b-scale-up-b}
        \includegraphics[width=0.32\linewidth]{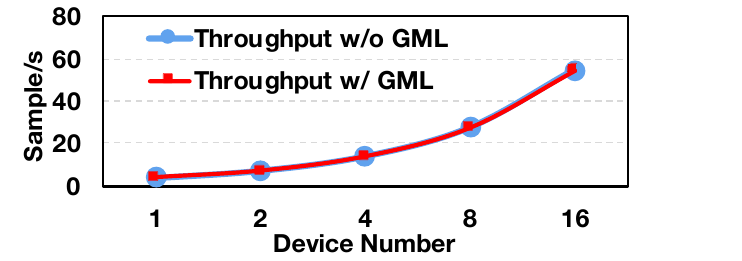}}
    \hfill
    \subfloat[GPT-NeoX-20B.]{\label{fig:GPT-NeoX-20b-scale-up-b}
            \includegraphics[width=0.32\linewidth]{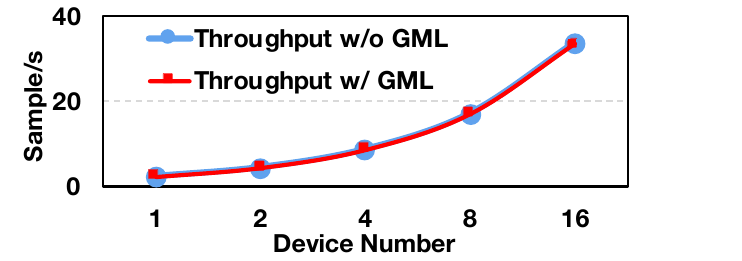}}
    \caption{Comparison of memory utilization ratio on GPU scale-out.}
    \label{fig:eval:gpuscaleup}
\end{figure*}

\begin{figure}
    \centering
        \includegraphics[width=0.4\textwidth]{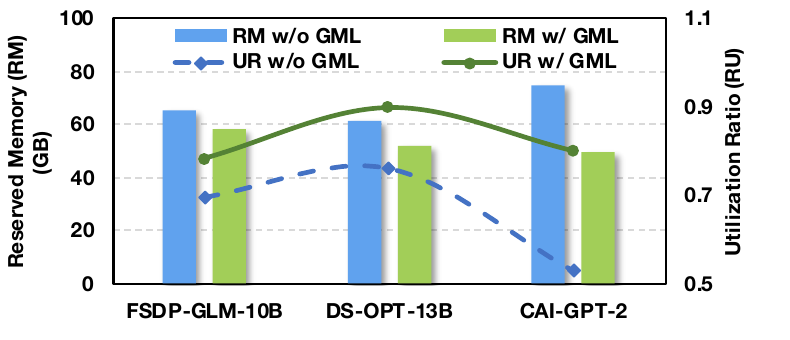}
    \caption{Memory utilization ratio comparison on various platforms.}
    \label{fig:eval:platforms}
\end{figure}

\subsection{Scalability of \proj{}}
Firstly, we conduct serveral experiments to evaluate the scalability of \proj{} across distinct memory-efficient strategies, GPU scale-out scenarios, and optimization platforms to address our observations in \Sec{sec:background}.

\subsubsection{Scalability on Memory-efficient Strategy}

To explore the scalability of \proj{} in terms of memory-efficient strategies, we conduct finetuning experiments on OPT-13B~\cite{zhang2022opt}, Vicuna-13B~\cite{vicuna2023}, and GPT-NeoX-20B~\cite{black2022gptneox20b} models using Deepspeed Zero3~\cite{DeepSpeed2023} with four NVIDIA A100 (80~GB) GPUs, all under a common batch size. Notably, our evaluation entails influential memory-efficient strategies, including LoRA, gradient-checkpointing (recomputation), and offload. Thus, we systematically employ combinations of these strategies during our assessment.
We label the no strategy scenario as \textbf{N}, recomputation as \textbf{R}, recomputation coupled with LoRA as \textbf{LR}, recomputation with offload as \textbf{RO}, and the joint utilization of recomputation, LoRA, and offload as \textbf{LRO}.

\begin{figure*}[t]
    \centering
    \subfloat[OPT-1.3B]{\label{fig:opt-1.3b-a}
        \includegraphics[width=0.32\linewidth]{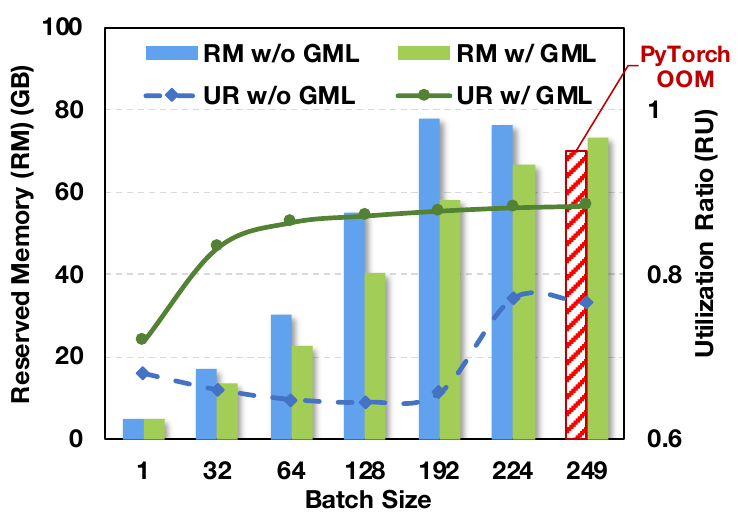}}
    \hfill
    \subfloat[OPT-13B]{\label{fig:opt-13b-a}
        \includegraphics[width=0.32\linewidth]{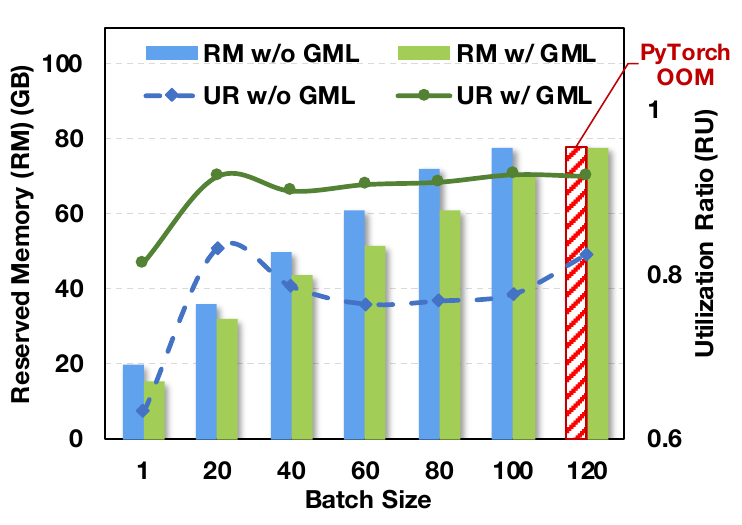}}
    \hfill
    \subfloat[GPT-NeoX-20B]{\label{fig:eval:end:gpt20b}
        \includegraphics[width=0.32\linewidth]{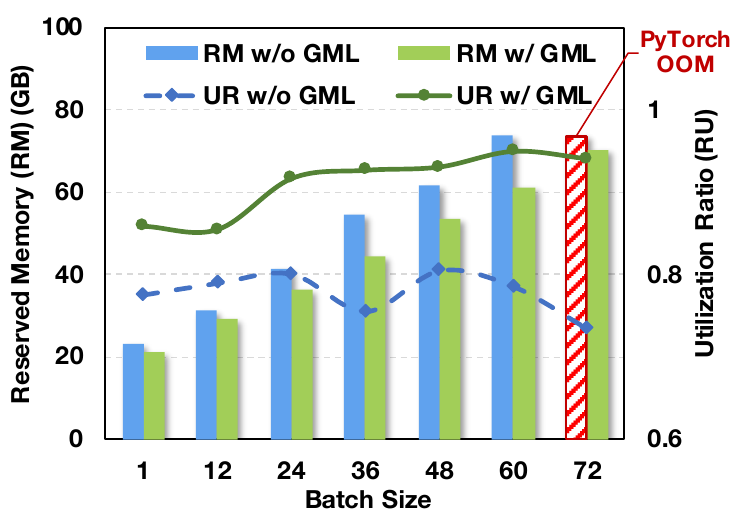}}
        \vspace*{4mm}
        \\
    \subfloat[OPT-1.3B]{\label{fig:opt-1.3b-b}
        \includegraphics[width=0.32\linewidth]{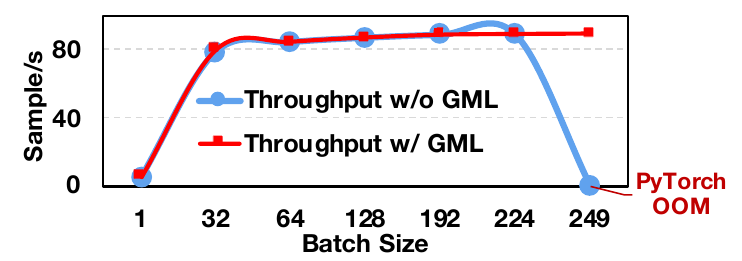}}
    \hfill
    \subfloat[OPT-13B]{\label{fig:opt-13b-b}
        \includegraphics[width=0.32\linewidth]{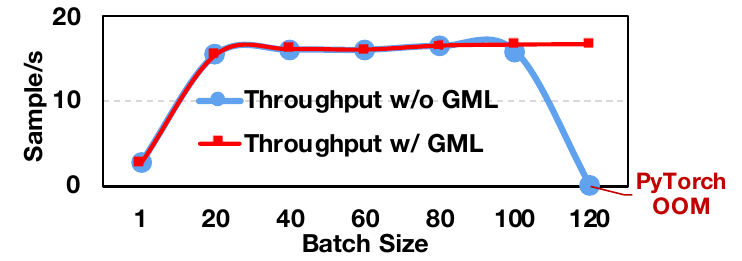}}
    \hfill
    \subfloat[GPT-NeoX-20B]{\label{fig:GPT-NeoX-20b-ZLR-b}
        \includegraphics[width=0.32\linewidth]{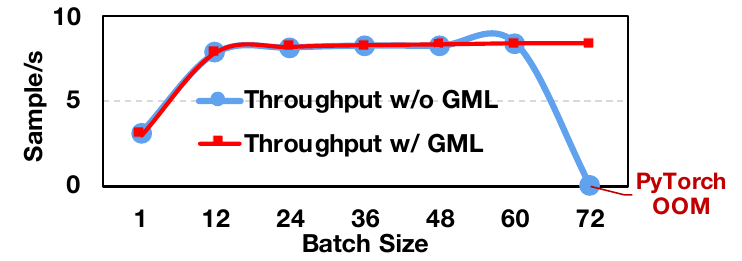}}
    \caption{Comparison of memory utilization ratio and throughput on end-to-end effectiveness, utilizing varying batch sizes.}\label{fig:eval:end}
\end{figure*}

\paragraph{Complicated strategies lead to fragmentation.} 
\Fig{fig:eval:strategy} illustrates the utilization ratio and reserved memory consumption. In a comprehensive overview, the increase in utilization ratio and reduction in reserved memory, when contrasted with PyTorch, ranges from approximately 5\% to 24\%  (or is around 10~GB and up to 17~GB), respectively. In contrast, \proj{} effectively reduce this fragmentation ratio to 5\% to 10\%, mitigating the fragmentation issue.

When complex optimization strategies are employed, the consequent fragmentation ratio on PyTorch can exceed 20\%. The contrast in utilization ratio between the application of these optimization strategies and their absence becomes evident. To illustrate, consider the recomputation strategy: it discards a portion of the activation tensor during the forward pass. 
This introduces a higher frequency of small memory allocation and deallocation operations, ultimately leading to fragmentation. A similar scenario arises with the offload strategy, where tensors frequently swap in and out between the CPU memory and the GPU memory, further increasing the frequency of memory allocation and deallocation operations. This situation highlights the need and features of \proj{}: its design is transparent and effective with those increasingly complex optimization strategies.

\subsubsection{Scalability on GPU Scale-out}
Employing LR strategies along with the Deepspeed platform, we proceed to evaluate \proj{}'s scalability in the context of GPU scale-out. This evaluation entails scaling from 1 GPU to 16 GPUs incrementally. As depicted in \Fig{fig:eval:gpuscaleup}, \proj{} consistently exhibits lower fragmentation ratios (high utilization ratio) and reserved memory consumption. Particularly noteworthy is the case of GPT-NeoX-20B in \Fig{fig:eval:gpuscaleup:gpt20b}, where the utilization ratio and reserved memory reduction can be as substantial as 23\% or 17~GB, respectively. Moreover, the trend becomes evident that, as the number of GPUs increases, \proj{} effectively maintains a utilization ratio of approximately 90\% (i.e., fragmentation ratio less than 10\%).

\paragraph{GPU scale-out leads to fragmentation.} 
While tremendous clusters speed up the training process, they also bring about more and more GPU memory fragmentation. \Fig{fig:eval:gpuscaleup} presents the utilization ratio gradually decreases when the GPU number scales up from 1 to 16. 
Despite certain outliers, the Figure shows that the fragmentation ratio is enlarged regarding PyTorch. 
To put it in detail, the distributed data-parallel strategy causes this trend. DeepSpeed ZeRO-3~\cite{rajbhandari2021zero} means partitioning the optimization states, gradients, and weights. When more GPU is involved in training, the above tensor size will be smaller. 
It will cause many large blocks to split and cause more memory defragmentation. 
As shown in \Fig{fig:eval:gpuscaleup} bottom, \proj{} also maintains high throughput, similar to the original PyTorch. 
That indicates \proj{} has excellent scalability with very low overhead.

\subsubsection{Scalability on Various Platforms}
We utilize various platforms including Deepspeed~\cite{rasley2020deepspeed}, FSDP~\cite{zhao2023pytorch}, and Colossal-AI~\cite{li2021colossal} to conduct finetuning on the OPT-13B~\cite{zhang2022opt}, GLM-10B~\cite{du2021glm}, and GPT-2~\cite{radford2019language} models, respectively. This process employs static optimizing strategies, specifically LoRA and recomputation, and involves the use of four NVIDIA A100 (80~GB) GPUs.
As illustrated in \Fig{fig:eval:platforms}, the results demonstrate a noteworthy decrease in both fragmentation and reserved memory, with reductions ranging from approximately 9\% to 33\%, and from 7~GB to 25~GB, respectively. These results confirm the high scalability exhibited by \proj{} on various optimized training platforms.

\subsection{End-to-End Effectiveness of \proj{}} \label{sec:effectiveness}
In this study, we conduct a comparative analysis between \proj{} and the PyTorch caching allocator through end-to-end fine-tuning of LLMs, utilizing varying batch sizes. This evaluation uses four A100 GPUs and enables LoRA, recomputation, and Zero3 optimizations on both frameworks. \Fig{fig:eval:end} shows that \proj{} consistently demonstrates a substantial reduction in peak memory consumption across a range of model sizes from 1.3 to 20 billion parameters. Notably, this memory consumption mitigation exhibits scalability with increasing model sizes while maintaining a consistent batch size.

The \fixme{efficiency} of memory usage is also demonstrated by the trends presented in Figure~\ref{fig:eval:end}. Significantly enhanced performance is evident in comparison to the baseline. Notably, as the model size increases, memory efficiency reaches levels exceeding 95\% (as seen with the 13 billion and 20 billion parameter models), indicating minimal fragmentation and waste. This starkly contrasts the baseline approach, which struggles to achieve an 80\% efficiency rate.

\fixme{\proj{} can reduce the framgmentaion and provide more memory to run LLM without facing OOM errors. Especialy, in {fig:eval:end}, we can see OPT-1.3B, OPT-13B and GPT-NeoX-20B perform well with \proj{} while encountering OOM errors with PyTorch's CUDA caching allocator in large batch scenarios, indicating effectiveness of \proj{}.}

Furthermore, we quantify the overhead of defragmentation logic overhead using the end-to-end throughput. 
Figure~\ref{fig:eval:end} bottom shows that \proj{} effectively maintains comparable throughput to the baseline approach. Interestingly, in scenarios involving exceedingly large batch sizes, \proj{} exhibits the potential to achieve even higher throughput than the PyTorch baseline due to its adept memory management and reduced frequency of (de)allocation operations.

\subsection{Memory Trace Analysis}
\label{sec:memtrace}
\begin{figure}[t]
    \begin{center}
    \includegraphics[width=0.48\textwidth]{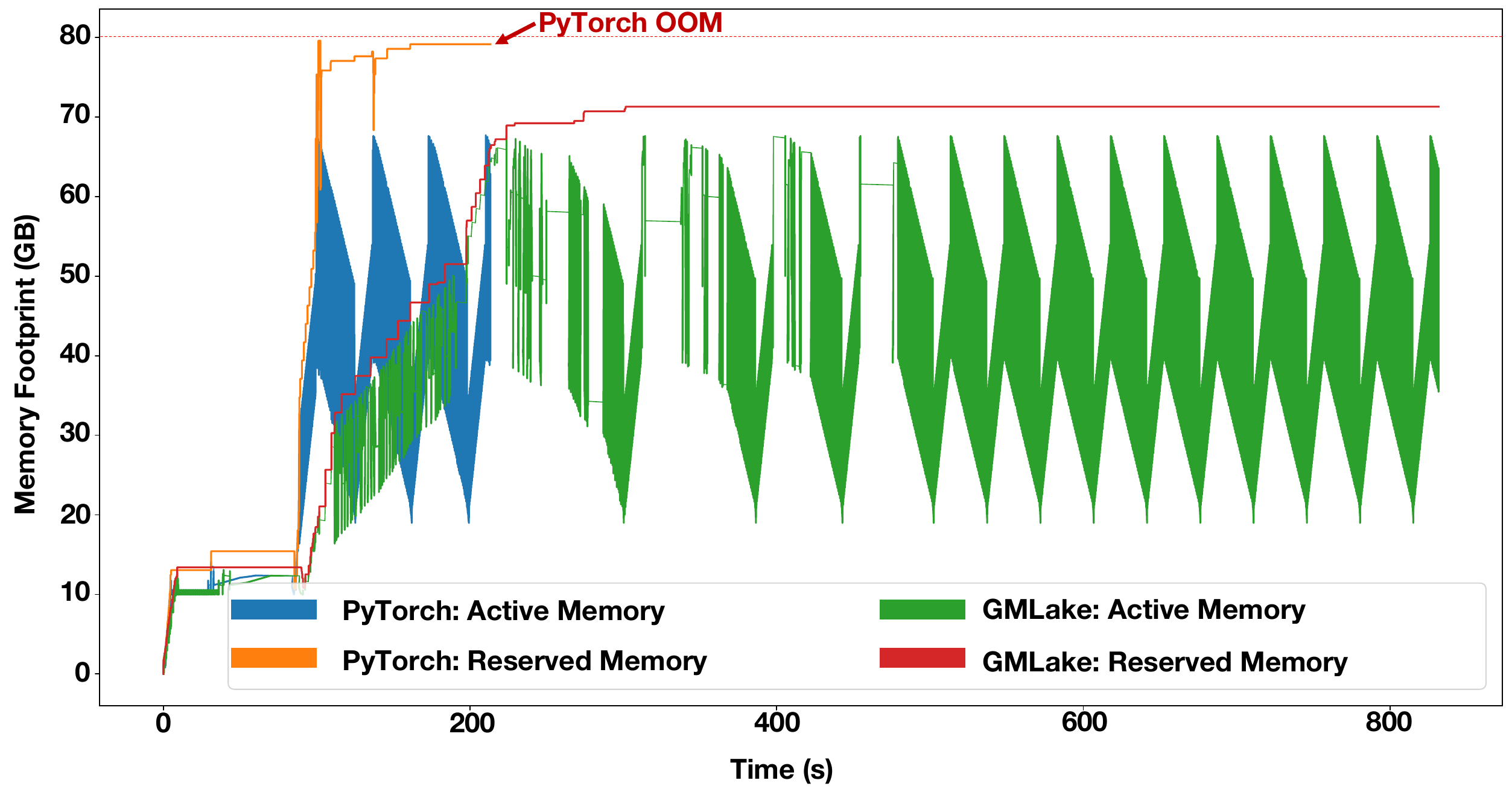}
    \caption{Memory trace on GPT-NeoX with batch size 72. }
    \label{fig:trace}
    \end{center}
\end{figure}

Finally, we track the memory allocation behaviors on GPT-NeoX-20B on 4 GPUs with LoRA and recomputation strategies for PyTorch and \proj{}, as shown in \Fig{fig:trace}.

There are three noteworthy points that highlights the advantages of \proj{}.
Firstly, PyTorch terminates at about 200~s owing to the OOM exception, while \proj{} functions correctly for this batch size.
Secondly, although the active memory of \proj{} and PyTorch is of the same level, their reserved memory differs greatly, indicating the large fragmentation issue in PyTorch.  

\fixme{
Thirdly, during the 100s to 400s, the active memory of both PyTorch and \proj{} fluctuates regularly, showing the memory pattern in the LLM fine-tuning stage, especially in forward and backward passes. The rise of active memory represents the forward pass memory allocation, while the decrease of active memory stands for the backward pass. The interval between two similar patterns reflects the time cost of one single iteration.
}
 
Lastly, after four iterations, \proj{} reaches stability and achieves the same throughput as PyTorch.
That indicates that the allocation strategy used in \proj{} can quickly adapt to the memory fluctuation in the forward/backward training passes, and find the best caching and stitching strategy.

\fixme{
We exploit the periodical nature of DNN training. \Fig{fig:trace} shows that DNN training has a stable period in the training, where similar VMM allocation requests repeat. This periodicity presents the opportunity to reuse stiched sBlocks, amortizing their cost. To achieve that, we design the stitched memory pool (sPool) in GMLake to overscribe the sBlocks. When a new sBlock is created, we add it to the sPool. When the sBlock is freed, we still keep it presence. Next time when the same sBlock needs to be created, it can directly reuse the previously created sBlock. As long as we maintain enough sPool instances, all allocations only search for its best-fit sBlock without creating a new sBlock. We call it convergence, e.g., after four iterations of \Fig{fig:trace}.
}

\section{Related Works and Discussion}\label{sec:relatedworks}
We compare \proj{} with existing works in two aspects: memory defragmentation and memory optimizations of LLMs. 

\paragraph{Memory Defragmentation}
The memory defragmentation has been extensively investigated in various contexts~\cite{FragProblem, MultiCoreLowFrag, EDDY}. In an effort to address fragment-related challenges, an early literature introduced a straightforward approach involving fine-grained fixed-sized chunks~\cite{FixedSize}. 
While this approach eliminates data movement overhead, it introduces increased access overhead and limited flexibility.
To enhance both efficiency and flexibility, researchers have proposed compaction-based strategies, such as those involving the consolidation of multiple small chunks into a larger, contiguous unit through data movement~\cite{ParallelDeOnGPU, ZGC}. 
Other defragmentation techniques, including copy-based garbage collection systems~\cite{ParallelCopyingGC, CopyingGCStopWorld, HCopyingGC}, reduce complexity in data movement logic at the expense of temporary memory wastage.
While sharing some conceptual similarities with the consolidation of small chunks into larger ones, \proj{} adopts a stitching-based technique, which minimizes the need for frequent data movement and copying, resulting in a significant enhancement of memory efficiency. Beyond conventional memory systems, recent research has also explored the defragmentation for persistent memories~\cite{FFCCD}.

\paragraph{Efficient LLM}
For Transformer-based LLMs, memory has emerged as a paramount resource within computing systems. 
The quadratic nature of attention mechanisms has led to a substantial surge in memory consumption for LLMs, thereby magnifying the significance of effective memory management~\cite{Bert, GPT3, GPT4}. 
Researchers have proposed various algorithmic optimizations aimed at curbing memory consumption, including quantization techniques~\cite{jacob2018quantization, wang2019learning, guo2020balancing, zhuang2021effective, li2022efficient, guo2022squant, ANT, Olive, AWQ}, pruning strategies~\cite{han2015deep,  zhu2019sparse, Qiu_2019_CVPR, guan2020far, qin2020sigma, guo2020accelerating,  wang2021dual, LLMPruner, Transkimmer, ZipLM}, and KV-cache compression approaches~\cite{Reformer, Unlimiformer}, compilation~\cite{TVM, Ansor, Roller, TASO, zhou2023ugrapher} and scheduling~\cite{LazyBatch, BubbleUp, BubbleFlux, Heracles, VELTAIR, guo2022nesting}.


There are various system-level memory optimizations.
The vLLM work leverages page-based virtual memory management techniques to substantially enhance resource efficiency and serving throughput~\cite{vLLM}. FlashAttention employs tiling techniques to optimize attention computation and notably mitigate memory consumption~\cite{FlashAttention, FlashAttention2}. The FlexGen framework introduces an optimization strategy to determine optimal memory-computation arrangements for efficient pipeline execution~\cite{FlexGen}. 
In this landscape, \proj{} emerges as a user-transparent memory management system that orchestrates intelligent and efficient memory reuse.

\paragraph{{Novelty against other  works}}
\label{sec:novelty}
{
\proj{} works on a unique memory scope for DNN training, which is different from the vLLM~\cite{vLLM} and vMalloc~\cite{virtualalloc2022}/CUDA VMM~\cite{perry2020introducing}. 
Neither them can solve the problem that the GMLake solved. }

{
The vLLM is an algorithm-based solution for the Self-Attention operation and acts inside a tensor. 
LLMs originally pad all sequences with variable-length tokens to the maximum length and cause significant redundancy.
vLLM uses a lookup table to remove the padded tokens. As such, vLLM is specific to self-attention while GMLake targets DNN training with a wider scenario.}
CUDA VMM, similar to vMalloc, is a low-level system tool for defragmentation. 
However, CUDA VMM/vMalloc cannot be aware of the memory pool commonly used in the DL framework. 
Without the memory pool design, the performance will significantly degrade. 
Therefore, these system-based memory tools cannot be directly implemented on the DL framework.

\begin{table}[t]
    \centering
        \resizebox{0.9\linewidth}{!}{
    \begin{tabular}{c|c}
    \toprule
    \textbf{Method} & \textbf{Scope}\\
    \midrule
    {vLLM~\cite{vLLM}} & Tensor\\
    { \proj{}} & Memory Pool\\
    { vMalloc/CUDA VMM~\cite{vicuna2023}} & Physical Memory\\
    \bottomrule
    \end{tabular}
    }
    \caption{Technical differences to related work.}
    \label{tbl:novelty}
\end{table}

\section{Conclusion}\label{sec:conclusion}

This study introduces \proj{}, a new memory allocator with high efficiency and low fragmentation.
Built upon the low-level CUDA virtual memory management knob,  it consolidates multiple non-contiguous physical memory blocks into a singular, contiguous entity, thus mitigating the fragmentation issue.ß
To facilitate seamless integration to existing DL frameworks, we formulate a virtual memory management API. Furthermore, we propose a multi-staged defragmentation strategies to guarantee the allocation efficiency and robustness.
Our evaluation shows that \proj{} reduces the memory fragmentation to 5\% $\sim$ 10\% while maintaining the same throughput.

\section*{Acknowledgements}

This work was supported by the National Key R\&D Program of China under Grant 2021ZD0110104, the National Natural Science Foundation of China (NSFC) grant (62222210, U21B2017, and 62072297), and the ANT Group Research Fund.
The authors would like to thank the anonymous reviewers for their constructive feedback for improving the work. 
Any opinions, findings, and conclusions in this paper are those of the authors only and do not necessarily reflect the views of our sponsors.

\bibliographystyle{plain}
\bibliography{references}

\end{document}